\documentclass{article}
\usepackage{epsfig}
\usepackage{amsmath}

\newcommand{\pt}{p_t}
\newcommand{\mt}{m_t}
\newcommand{\ct}{{\rm t}}
\newcommand{\cz}{{\rm z}}
\newcommand{\cx}{{\rm x}}
\newcommand{\cy}{{\rm y}}
\newcommand{\rt}{\tilde r}
\newcommand{\phb}{\phi_b}
\newcommand{\phs}{\phi_s}
\newcommand{\bkj}{\mbox{K}_1}
\newcommand{\bid}{\mbox{I}_2}
\newcommand{\bin}{\mbox{I}_0}

\newcommand{\bt}{\beta_t}

\begin{document}
%
%
\title{Disentangling Spatial and Flow Anisotropy}
\author{Boris Tom\'a\v sik \\
The Niels Bohr Institute, Blegdamsvej 17,\\ %
2100 Copenhagen \O, Denmark}
\date{February 17, 2005}
\maketitle
%
%
\begin{abstract}
Two generalisations of the blast-wave model to non-central 
nuclear collisions are
constructed, and elliptic flow as well as azimuthal dependence of 
correlation radii are calculated. Particular attention is paid to how 
different azimuthal dependences of transverse 
flow direction can cause qualitatively different anisotropic 
fireballs to give same $v_2$ as a function of the transverse momentum.
The simultaneous dependence of $v_2$ and the oscillation of correlation 
radii on both spatial and flow anisotropy is studied in great detail.
\end{abstract}
%

\section{Introduction}

The so-called ``elliptic flow'' $v_2$ \cite{vol}, observed in 
non-central nuclear collisions at highest available energies
\cite{phv2,stv2}, has had important implications on understanding
of the collision dynamics in framework of hydrodynamic and cascade
models \cite{Heinz:2001xi,Molnar:2001ux}. After exclusion of non-flow
effects the ``elliptic flow'' can be caused by azimuthal anisotropy 
in transverse expansion and/or anisotropic shape of the fireball. 
It was noticed that it was not possible to conclude on the 
spatial anisotropy of the freeze-out state simply from data on 
$v_2$. However, a conjecture \cite{st130} was made that measurements of HBT 
correlation radii as functions of azimuthal angle \cite{Wiedemann:1997cr}
will give access
to the spatial shape of the fireball. Indeed, this
was observed \cite{khplb} in hydrodynamic simulations with two different sets 
of initial conditions: they lead to different final states which could 
have been distinguished by pion interferometry. 

This paper focuses in great detail on the question to what 
extent the spatial shape {\em and} the anisotropy of transverse flow can 
be identified from data. In contrast to hydrodynamic simulations where 
given model and initial conditions lead to a single freeze-out state,
parameterisations of the final state will be employed here. These 
parameterisations, which will be constructed by generalising the 
blast-wave model \cite{Csorgo:1995bi}, allow to investigate a broad range 
of various freeze-out states and find their possible signatures
in the data. A much more systematic study than in a hydrodynamic simulation
is thus possible; the price to pay is that no connection to fireball
evolution is made.

The generalisation of the blast-wave model to non-central collisions
is not unique. In order to explore possible ambiguities due to 
various angular dependences of the expansion velocity, I will study 
two models.

There is a correlation between the spatial 
and the flow anisotropy in determining $v_2$; same $v_2$ can be
caused by many different combinations of the two anisotropies.
This correlation depends on the mass of particles 
but---unfortunately---it 
also depends crucially on the used model. Thus the 
flow anisotropy cannot be disentangled from the spatial anisotropy 
unless the model is known. On the other hand, irrespective the model,
the azimuthal dependence 
of correlation radii seems to be mostly sensitive to the spatial 
anisotropy, at least in the low-$\pt$ region. 

First, in the next Section I will introduce a generalised blast-wave model.
Then, $v_2$ (Section \ref{v2})  and correlation radii (Section \ref{s:cr})
are calculated and their dependence on spatial and flow anisotropy
is studied.
Technical details concerning the calculations can be found in 
Appendices.


\section{A generalisation of the blast-wave model}
\label{model}

This generalisation follows mainly ref.~\cite{retiere} with 
some variations in introducing the azimuthal dependence of the 
transverse velocity. (Experts in the field can skip most of this
section and look just at the introduction of two different azimuthal 
dependences of the transverse flow velocity, after eq.~\eqref{vprho}.)

It is assumed that at the end of its evolution the fireball is 
in a state of {\em local thermal equilibrium} characterised by
a temperature $T$. Decoupling of particles is (almost) instantaneous 
and can be modelled by the Cooper-Frye formalism \cite{cf} along a 
freeze-out hyper-surface. 

The time of freeze-out does not depend on position in direction 
transverse to the beam, only the longitudinal coordinate matters. 
Motivated by the Bjorken boost-invariant longitudinal expansion
\cite{bjork}, 
the freeze-out hyper-surface is given by a hyperbola
\begin{equation}
\tau_0 = \sqrt{\ct^2 - \cz^2} = \mbox{const}\, ,
\end{equation}
where $\ct$ and $\cz$ are temporal and longitudinal coordinate, respectively.
We will allow for some smearing of the freeze-out time $\tau_0$
by amount $\Delta\tau$. 

We assume that the decoupling matter is distributed uniformly and the 
transverse cross-section has ellipsoidal shape. Thus the density will 
be proportional to $\Theta(1 - \rt)$ where 
\begin{equation}
\label{rtil}
\rt = \sqrt{\frac{\cx^2}{R_x^2} + \frac{\cy^2}{R_y^2}}\, .
\end{equation}
In this equation, $\cx$ and $\cy$ are Cartesian coordinates in the 
direction of the impact parameter and perpendicular to the reaction plane,
respectively. They can be rewritten with the help of the usual 
radial coordinates
\begin{subequations}
\label{spc}
\begin{eqnarray}
\cx & = & r\, \cos\phs \, ,\\
\cy & = & r\, \sin\phs \, ,
\end{eqnarray}
\end{subequations}
(the reason for subscript ``$s$'' on the angular coordinate will 
become clear later).
The radii $R_x$ and $R_y$ in eq.~\eqref{rtil} stand for the sizes in the
corresponding directions. They will be expressed via the average radius 
$R$ and the spatial anisotropy parameter $a$
\begin{equation}
\label{adef}
R_x = a\, R\, , \qquad R_y = R/a\, .
\end{equation}
Thus a fireball elongated out of the reaction plane corresponds  
to $a<1$, while $a>1$ stands for an in-plane elongated source.

We do not assume any geometric limitation in the longitudinal
direction, therefore the fireball is actually infinite in this
direction. We can do this as we will be only interested in observables 
at mid-rapidity in collisions at very high energy (at RHIC e.g.)
where boost invariance is locally established. The actual finiteness
of the effective source is established dynamically \cite{Makhlin:1987gm}. 
Those parts
of the fireball moving too fast forward or backward cannot
emit mid-rapidity particles.

The source is modelled by an emission function. This is the 
Wigner phase space density of particle emission
\begin{multline}
\label{efun}
S(x,p)\, d^4x = \frac{\mt\, \cosh(y-\eta)}{(2\pi)^3}\, 
d\eta\, d\cx\, d\cy\, \frac{\tau\, d\tau}{\sqrt{2\pi\, \Delta\tau}} \\
\times
\exp\left( -\frac{(\tau - \tau_0)^2}{2\, \Delta\tau^2}\right ) \,
\Theta(1-\rt) \, \exp\left ( -\frac{p^\mu u_\mu}{T} \right )\, .
\end{multline}
%
\begin{table}[t]
\caption{Summary of model parameters.
\label{t:params}}
\begin{center}
\begin{tabular}{lc}
freeze-out temperature               & $T$           \\
average transverse flow gradient     & $\rho_0$      \\
variation of the flow gradient       & $\rho_2$      \\
average transverse radius            & $R$           \\
spatial anisotropy                   & $a$           \\
mean Bjorken lifetime                & $\tau_0$      \\
freeze-out time dispersion           & $\Delta\tau$  
\end{tabular}
\end{center}
\end{table}
%
Here, we use space-time rapidity $\eta$ and longitudinal proper time
$\tau$ instead of $\ct$ and $\cz$
\begin{subequations}
\begin{eqnarray}
\ct  & = &  \tau\, \cosh\eta\, ,\\
\cz  & = &  \tau\, \sinh\eta\, .
\end{eqnarray}
\end{subequations}
Momentum $p$ will be parametrised in terms of rapidity $y$,
transverse momentum $\pt$, transverse mass $\mt = \sqrt{m^2 + \pt^2}$
and azimuthal angle $\phi$
\begin{equation}
\label{pdec}
p^\mu = (\mt\cosh y, \, \pt\cos\phi,\, \pt\sin\phi,\, \mt\sinh y)\, .
\end{equation}
The term $\mt\cosh(y-\eta)$ in the emission function comes from 
the flux of particles through an infinitesimal piece of the freeze-out
hyperbola: $p^\mu d\sigma_\mu$ \cite{cf}. In the Boltzmann distribution,
energy is taken in the rest frame of the emitting piece of the fireball
\begin{equation}
E^* = p^\mu u_\mu(x)\, ,
\end{equation}
where $u_\mu$ is local collective velocity of the fireball. The use
of Boltzmann distribution is justified as long as the temperature is
not too low and the chemical potential 
(for pions) is small; here we put $\mu = 0$.

Velocity field $u_\mu(x)$ describes the collective expansion of 
the fireball \cite{Csorgo:1995bi}. 
In longitudinal direction we assume a boost-invariant
expansion which is given by
\begin{equation}
v_z = \tanh\eta\, .
\end{equation}
The transverse velocity will be parametrised with the help of 
transverse rapidity $\rho$
\begin{equation}
v_\perp = \tanh\rho\, .
\label{vprho}
\end{equation}
Rapidity $\rho$ will depend on the position in  the transverse plane.
We will consider two models which will differ in the azimuthal variation 
of the transverse velocity.

\textbf{Model 1.} 
In this model transverse expansion velocity is always directed 
perpendicularly to the surface given by $\rt = \mbox{const}$ \cite{retiere}.
Its angle with respect to the reaction plane is thus 
\begin{equation}
\phb = \mbox{Arctan}\frac{\cy}{\cx}
\end{equation}
%
\begin{figure}[t]
\begin{center}
\begin{minipage}{5.6cm}
\centerline{\epsfig{file=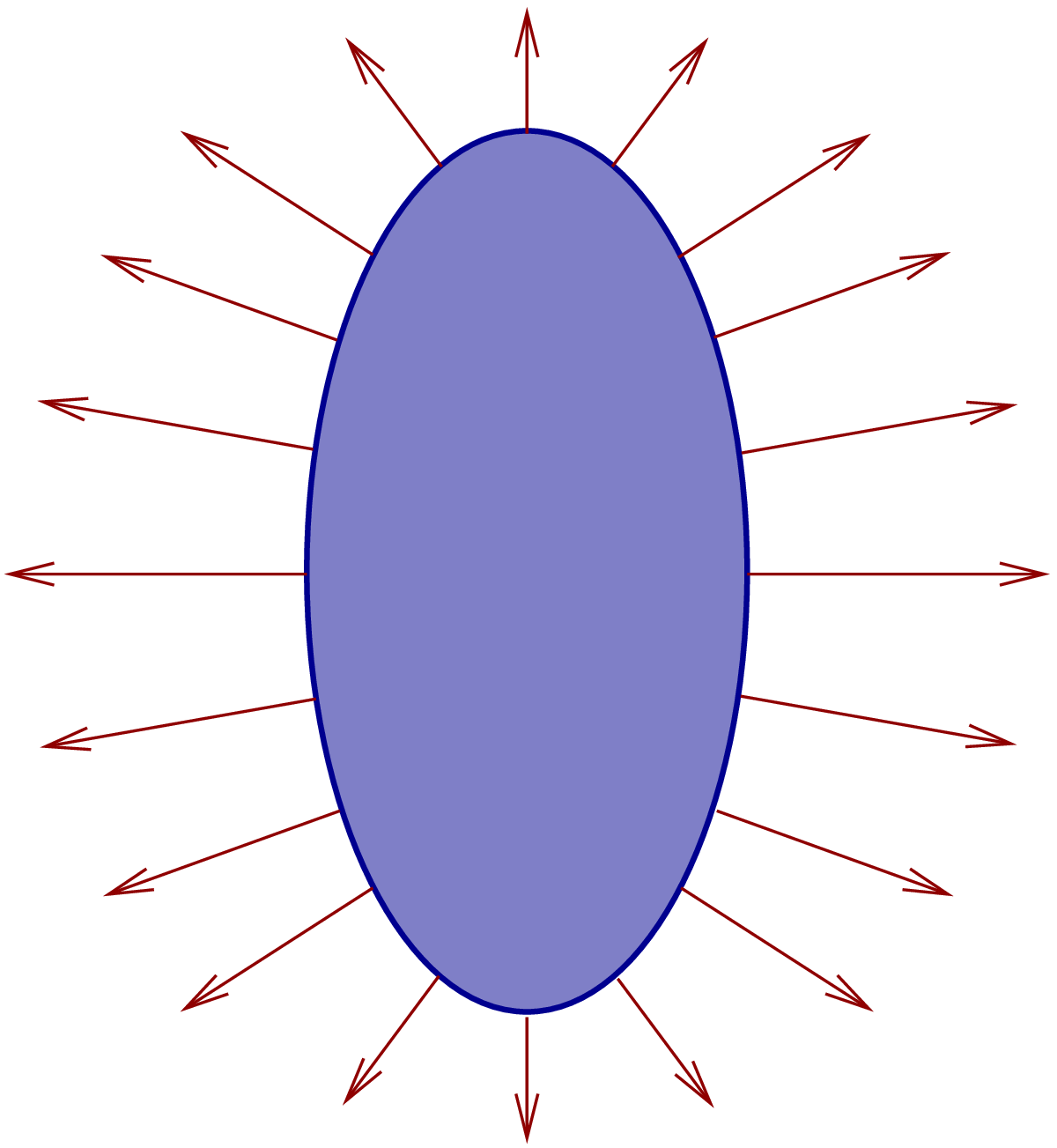,width=5.5cm}}
\centerline{Model 1}
\end{minipage}
\begin{minipage}{5.6cm}
\centerline{\epsfig{file=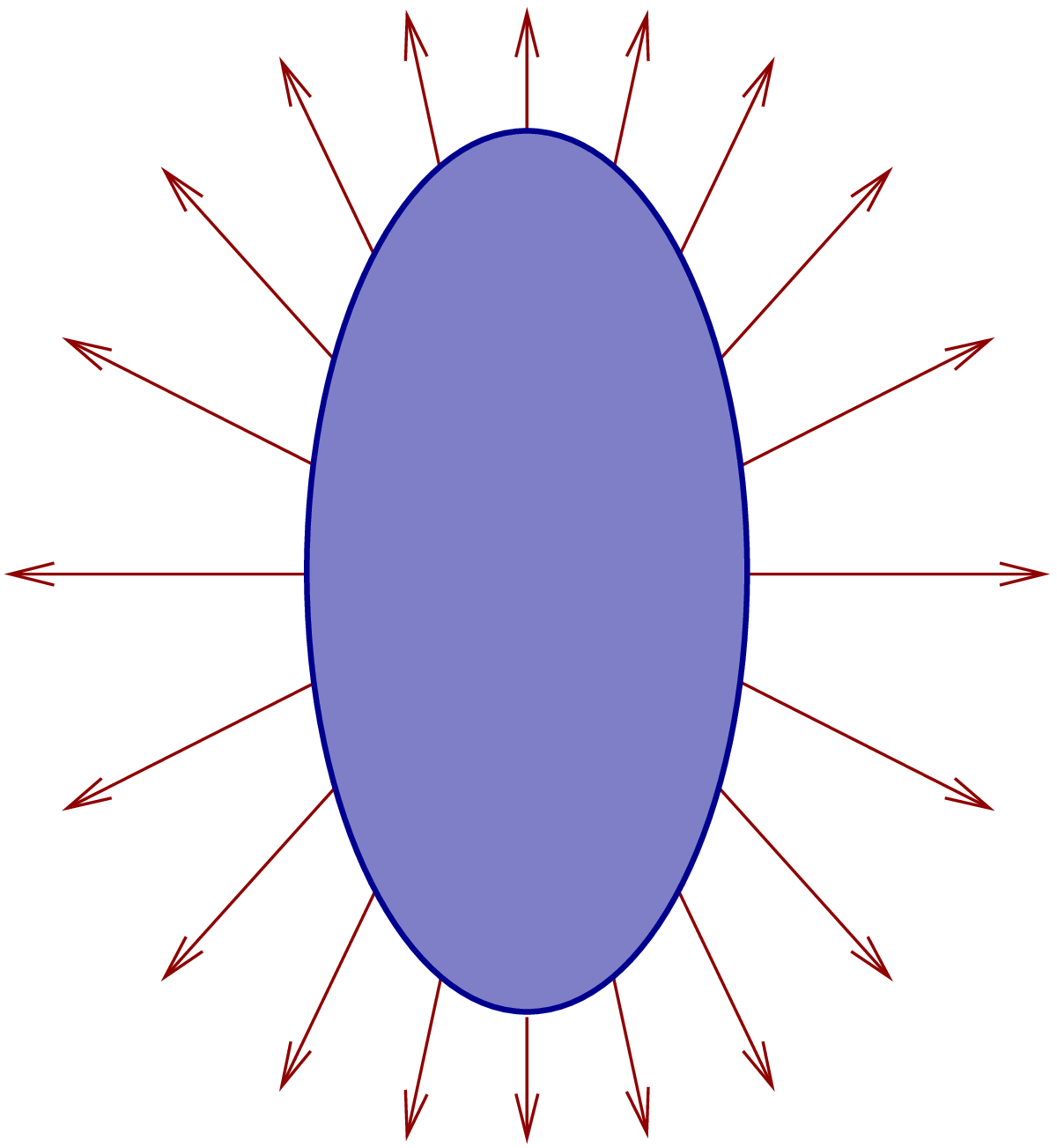,width=5.5cm}}
\centerline{Model 2}
\end{minipage}
\end{center}
\caption{%
Two used models of azimuthal variation of the transverse flow
velocity. The depicted anisotropies correspond to $a<1$ and 
$\rho_2 > 0$.
\label{f:models}}
\end{figure}
(see Figure \ref{f:models}). Note that 
\begin{equation}
\label{pbps}
\tan\phb = \left ( \frac{R_x}{R_y}\right )^2 \tan\phs = a^4\tan\phs
\, .
\end{equation}
The magnitude of the transverse rapidity also varies with  $\phb$
\begin{equation}
\label{rhom1}
\rho = \rt\, \rho_0 (1 + \rho_2\cos(2\phb))\, ,
\end{equation}
where $\rho_0$ and $\rho_2$ are tunable parameters.
(Note the slight difference to the parametrisation of Reti\`ere and Lisa
\cite{retiere} who write $\rho = \rt  (\rho_0 + \rho_2\cos(2\phb))$.)

The velocity field $u^\mu (x)$ is then written as
\begin{equation}
\label{um1}
u^\mu = (\cosh\rho\, \cosh\eta,\, \cos\phb\, \sinh\rho,\, 
         \sin\phb\, \sinh\rho,\, \cosh\rho\, \sinh\eta)\, .
\end{equation}

Later in the calculation it will be convenient to use coordinates
$\rt$ and $\phb$ in the transverse plane instead of $\cx$ and $\cy$.
It is shown in Appendix~\ref{sjaco} that 
\begin{subequations}
\label{jac1}
\begin{eqnarray}
d\cx\, d\cy & = & J_1(\phb)\, R^2\, \rt\, d\rt\, d\phb\, , \\
J_1(\phb)   &  =  & (a^2\cos^2\phb + a^{-2}\sin^2\phb)^{-1} \, .
\end{eqnarray}
\end{subequations}

\textbf{Model 2.}
Here it will be assumed that the transverse velocity is always directed 
radially. Then the angle between transverse velocity and the reaction 
plane coincides with $\phs$ (see eq.~\eqref{spc}). Transverse rapidity 
will be given by
\begin{equation}
\label{rhom2}
\rho = \rt\, \rho_0 (1 + \rho_2 \cos(2\phs))\, .
\end{equation}
The difference to Model 1 is in the use of $\phs$ instead of 
$\phb$, cf. equation \eqref{rhom1}. The velocity field 
is similar to eq.~\eqref{um1}, except for the replacement
$\phb\to\phs$
\begin{equation}
\label{um2}
u^\mu = (\cosh\rho\, \cosh\eta,\, \cos\phs\, \sinh\rho,\, 
         \sin\phs\, \sinh\rho,\, \cosh\rho\, \sinh\eta)\, .
\end{equation}

In this case, the appropriate coordinates to use are $\rt$
and $\phs$. The Jacobian is calculated in Appendix~\ref{sjaco}
\begin{subequations}
\label{jac2}
\begin{eqnarray}
d\cx\, d\cy & = & J_2(\phs)\, R^2\, \rt\, d\rt\, d\phs \\
J_2(\phs)   & = & (a^{-2} \cos^2\phs + a^2\sin^2 \phs )^{-1}\, .
\end{eqnarray}
\end{subequations}

All parameters of the models are summarised in Table~\ref{t:params}.


\section{Elliptic flow}
\label{v2}

The elliptic flow coefficient $v_2$ is introduced through the 
Fourier decomposition of the azimuthal dependence of single-particle
spectrum \cite{vol}. 
At mid-rapidity in symmetric collision systems such a
decomposition includes only even cosine terms
\begin{eqnarray}
\nonumber
P_1(\pt,\phi) & = & \left .\frac{d^3N}{\pt\, d\pt\, dy\, d\phi} \right |_{y=0} 
\\ & = & 
\frac{1}{2\pi}\, \left . \frac{d^2N}{\pt\, d\pt\, dy} \right |_{y=0}\, 
\left ( 1 + 2 v_2(\pt)\cos(2\phi) + \dots \right )\, .
\end{eqnarray}
In this formulation, $\phi$ is the angle between the transverse 
momentum and the reaction plane. The coefficient $v_2$ can thus be 
calculated as
\begin{equation}
\label{v2comp}
v_2(\pt) = \frac{\int_0^{2\pi} P_1(\pt,\phi)\, \cos(2\phi)\, d\phi}%
{\int_0^{2\pi} P_1(\pt,\phi)\, d\phi}\, .
\end{equation}
Single-particle spectrum is obtained from the emission function 
by integrating over the space-time
\begin{equation}
\label{ps}
P_1(\pt,\phi) = \int d^4x\, S(x,p)\, .
\end{equation}
Combining eqs.~\eqref{v2comp} and \eqref{ps} we obtain expressions for 
$v_2$ in our models; see Appendix~\ref{v2calc} for details of the calculation.
For Model 1 (velocity perpendicular to the surface) we obtain
\begin{multline}
\label{v21}
v_2 = \frac%
{\int_0^1  d\rt\, \rt \int_0^{2\pi}d\phb\, \cos(2\phb)\, J_1(\phb)\,
\bkj \left (\frac{\mt\cosh\rho(\rt,\phb)}{T} \right)
\bid \left (\frac{\pt\sinh\rho(\rt,\phb)}{T} \right)}%
{\int_0^1  d\rt\, \rt \int_0^{2\pi}d\phb\, J_1(\phb)\,
\bkj \left (\frac{\mt\cosh\rho(\rt,\phb)}{T} \right)
\bin \left (\frac{\pt\sinh\rho(\rt,\phb)}{T} \right)}\, \\
\mbox{[Model 1]} 
\end{multline}
while for Model 2 (radially directed  transverse velocity) we have
\begin{multline}
\label{v22}
v_2 = \frac%
{\int_0^1  d\rt\, \rt \int_0^{2\pi}d\phs\, \cos(2\phs)\, J_2(\phs)\,
\bkj \left (\frac{\mt\cosh\rho(\rt,\phs)}{T} \right)
\bid \left (\frac{\pt\sinh\rho(\rt,\phs)}{T} \right)}%
{\int_0^1  d\rt\, \rt \int_0^{2\pi}d\phs\, J_2(\phs)\,
\bkj \left (\frac{\mt\cosh\rho(\rt,\phs)}{T} \right)
\bin \left (\frac{\pt\sinh\rho(\rt,\phs)}{T} \right)}\, \\
\mbox{[Model 2]}
\end{multline}
where $\bkj$, $\bin$, and $\bid$ are modified Bessel functions.
Since we integrate over the angle, the only difference between
the two results is in the use of the Jacobian terms $J_1$ or
$J_2$. Moreover, the difference between these two terms is only 
in the replacement $a\to a^{-1}$. As the anisotropy parameter $a$ 
does not appear anywhere else in relations \eqref{v21} and \eqref{v22},
any $v_2$ calculated in one model is equal to $v_2$ calculated 
in the other model under transformation $a\to a^{-1}$.
Thus we have an {\em analytic example of two models which lead to the 
same $v_2$, while one is elongated in-plane and the other out-of-plane}.
This clearly demonstrates that {\em there is no possibility
to distinguish in-plane source from out-of-plane source just by 
measuring $v_2$.}

Physics reason behind the observation that the two models 
are ``inverse'' to each other can be deduced from Figure~\ref{f:models}.
The arrows denote  expansion velocity and their lengths indicate 
its magnitude. Both situations in that Figure correspond to 
$\rho_2>0$ and $a<1$. In case of Model 1, most arrows point rather 
in the reaction plane (which is taken to be horizontal in this Figure),
so the major boost effect leading to enhancement of the spectrum 
happens in this direction. For Model 2, a larger part of the flow
is directed out of the reaction plane and the enhancement of 
spectra due to the boost is in that direction. Thus Model 1
would lead to positive $v_2$ while Model 2---in this setup---to
negative $v_2$. This is, of course, just a qualitative argument. 
The equivalence of the two models under replacement $a\to a^{-1}$
is derived analytically.

We can therefore calculate $v_2$ just for one of the models;
results for the second one are then obtained trivially. I will 
choose Model 1.

Dependence of $v_2$ on the transverse momentum and particle identity
in Model 1 was thoroughly studied in \cite{retiere}. In the data 
\cite{phv2,stv2} for small $\pt$, $v_2$ is  positive, 
increases with increasing $\pt$ and decreases with growing 
mass of particles. This behaviour is reproduced in Model 1 if
$a<1$ and $\rho_2>0$. (Loosely speaking, one of these conditions
may be broken, but not ``too much''; see later when the results are 
shown.) In this parameter region, calculation shows that $v_2$ 
of heavier particles can become negative at low $\pt$ and start growing 
and be positive above some value of $\pt$ \cite{retiere,huo}. Such a dip to 
negative $v_2$ is also observed by STAR Collaboration for antiprotons 
in certain centrality bins  \cite{stv2}, but the effect may not be 
statistically significant.

We will be interested in how the spatial and the flow anisotropies 
are entangled in determining $v_2$ for various identified particle 
species.

In comparing $v_2$ of different species it turns out to be
unwise to use the $\pt$-averaged $v_2$. This is because the averages 
are weighted with the single-particle spectra which are not alike 
for different species: those for heavier particles are flatter. 
Hence, averaging $v_2(\pt)$ for heavy particles can sometimes
lead to larger resulting values than the same procedure yields with 
light particles, although the value of $v_2$ at any $\pt$ is lower 
for heavy particles. The reason is that flatter spectrum for heavy
particles gives stronger weight to larger $v_2$ at higher $\pt$.

%
\begin{figure}
\centerline{\epsfig{file=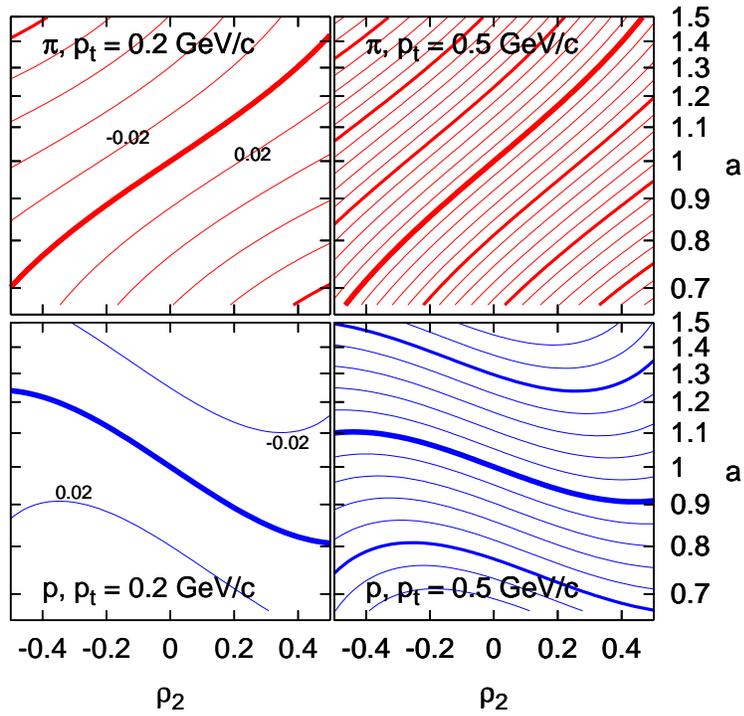,width=15.1cm}}%
\vspace*{-1.1cm}
\caption{%
Contour plots of 
$v_2$ as a function of $a$ and $\rho_2$, calculated in Model 1
for $T=100\, \mbox{MeV}$ and $\rho_0=0.88$. Upper row: pions, 
lower row: protons. Left column: $\pt = 200\, \mbox{MeV}/c$,
right column: $\pt = 500\, \mbox{MeV}/c$. The thickest lines
show where $v_2$ vanishes. Consecutive lines correspond to 
increments/decrements by 0.02.
\label{f:v2}}
\end{figure}
%
Therefore, we study the entanglement of $a$ and $\rho_2$ in determining 
$v_2$ for pions and protons at two fixed values of $\pt$. As can 
be seen from eqs.~\eqref{v21} and \eqref{v22}, $v_2$ does not 
depend on $R$, $\tau_0$ and $\Delta\tau$. Since the dependence
on all other parameters was studied in detail in \cite{retiere},
here we just fix $T$ and $\rho_0$, and plot $v_2$ as a function 
of $a$ and $\rho_2$ in Figure~\ref{f:v2}. We clearly see how 
the two anisotropy parameters are correlated and that the 
correlation depends strongly on the type of particles. Recall 
that a figure for Model 2 would be obtained just by inverting
the $a$-scale.

Temperature and radial flow can be roughly obtained from azimuthally 
integrated spectra which are weakly sensitive to $a$ and $\rho_2$
\cite{retiere}.
Then, {\em if one was able to determine the correct model} for the 
description of the freeze-out, {\em spatial and flow anisotropy could be 
disentangled from measurements of $v_2$ as a function of $\pt$ 
for different identified particle species} \cite{stv2}. However, the choice
of the model cannot be based on $v_2$ measurement.


\section{Azimuthal angle dependence of correlation radii}
\label{s:cr}

In this section we focus on azimuthal angle dependence of correlation 
radii due to spatial and flow anisotropy. First, explicit and 
implicit $\phi$-dependences are discussed. Then, necessary 
formalism is introduced. Experts can skip this and proceed directly 
to Section~\ref{hbt:results} where the results are presented.

\subsection{Explicit and implicit azimuthal angle dependence}

In non-central collisions one can study Bose-Einstein 
correlations of identical pions for particles emitted 
under different azimuthal angles \cite{Wiedemann:1997cr}.
To start the argument, let us just take a single emitter which 
emits particles in all directions. 
The directions in which the sizes of the emitter  
are measured, are given by the momentum. Therefore, 
by changing azimuthal angle $\phi$ of the momentum, the correlation 
radii measure the size of the 
anisotropic source in different directions. This 
leads to {\em explicit} dependence of the correlation radii
on $\phi$. (See below how correlation radii are defined.) 
The explicit azimuthal dependence is thus connected with the 
spatial anisotropy.

In a real case, we have an expanding fireball of which only a 
part---the {\em homogeneity region}---effectively produces particles
with a given momentum \cite{Makhlin:1987gm}. Thus particles in
different directions can be produced from different homogeneity 
regions which differ in sizes. This mechanism leads to an additional,
so-called {\em implicit} azimuthal dependence of the correlation 
radii. It is intimately connected with transverse expansion and
its anisotropy. Here we want to see how these two kinds of effects 
act together in azimuthally sensitive correlation studies.

\subsection{Formalism}

We will confine ourselves to theoretical calculations at mid-rapidity
for symmetric collision systems. The reader is referred 
to \cite{Tomasik:2002rx,Heinz:2002au} for summary of the formalism of 
Bose-Einstein interferometry in non-central collisions.

Correlation radii are width parameters of a Gaussian parametrisation 
of the measured correlation function
\begin{multline}
C(q,K) = 1+ \lambda\, \exp (-R_s^2(K)q_s^2 - R_o^2(K)q_o^2 - R_l^2(K)q_l^2 \\
{} - 2 R_{os}^2(K)q_o q_s - 2 R_{ol}^2(K) q_o q_l 
- 2 R_{sl}^2(K) q_s q_l )\, 
\end{multline}
where the momenta of the pair have been parametrised in terms of
\begin{subequations}
\begin{eqnarray}
q & = & p_1 - p_2 \\
K & = & \textstyle{\frac{1}{2}} (p_1 + p_2)
\end{eqnarray}
\end{subequations}
and the phenomenological parameter $\lambda \le 1$ is due to a variety 
of effects ranging from partial coherence of the source up till 
particle misidentification. The standard {\em out-side-long} coordinate
system is used, with longitudinal axis in beam direction, outward axis
parallel to the transverse component of $K$, and sideward direction 
perpendicular to the previous two. Correlation radii are given by sizes
in these directions. Recall that in non-central collisions we identify
the Cartesian x-y-z frame with the collision geometry: $\cz$-axis points 
in beam direction, $\cx$-axis is parallel
to the impact parameter, and $\cy$-axis is perpendicular to the reaction 
plane. Hence, there is an angle $\phi$ between the $\cx$-axis and the outward 
direction and we are interested in the $\phi$-dependence of the 
correlation radii.

If there is no tilt of the fireball in the reaction plane \cite{Lisa:2000xj},
the two radii $R_{ol}^2$ and $R_{sl}^2$ vanish. This is the case with the 
used models. The remaining radii can be calculated as 
\cite{Wiedemann:1997cr}
\begin{subequations}
\label{mix}
\begin{eqnarray}
R_s^2 & = & \textstyle{\frac{1}{2}} 
             (\langle\tilde\cx^2\rangle + \langle \tilde\cy^2\rangle )
             - \textstyle{\frac{1}{2}}
             (\langle \tilde{\cy}^2\rangle - \langle \tilde{\cx}^2\rangle)
             \cos 2\phi 
             - \langle \tilde{\cx}\tilde{\cy}\rangle 
                       \sin 2\phi \, ,
                      \label{rsmi} \\
R_o^2 & = & \textstyle{\frac{1}{2}} 
             (\langle\tilde\cx^2\rangle + \langle \tilde\cy^2\rangle )
             + \textstyle{\frac{1}{2}}
             (\langle \tilde{\cy}^2\rangle - \langle \tilde{\cx}^2\rangle)
             \cos 2\phi 
               + \langle \tilde{\cx}\tilde{\cy}\rangle \sin 2\phi 
                      \nonumber \\
                      & & {} + \beta_\perp^2 \langle \tilde{\ct}^2\rangle\, 
                      - 2\bt
                       \langle \tilde{\ct}\tilde{\cx} \rangle \cos\phi 
                     - 2\bt 
                       \langle \tilde{\ct} \tilde{\cy} \rangle \sin\phi\, , 
                      \label{romi} \\
R_{os}^2 & = & \textstyle{\frac{1}{2}}  
                  (\langle \tilde{\cy}^2\rangle - \langle \tilde{\cx}^2\rangle)
                  \sin 2\phi
                  + \langle \tilde{\cx}\tilde{\cy}\rangle \cos 2\phi
                  \nonumber \\
                  && {} + \bt \langle \tilde{\ct}
                       \tilde{\cx}\rangle \sin\phi
                     - \bt \langle \tilde{\ct}
                       \tilde{\cy}\rangle \cos\phi \, ,
                      \label{rosmi} \\
R_{l}^2 & = & \langle (\tilde{\cz} -\beta_l\tilde{\ct})^2 \rangle \, ,
                      \label{rlmi} 
\end{eqnarray}
\end{subequations}
where
\begin{subequations}
\begin{eqnarray}
\langle f(x) \rangle(K) & = & 
\frac{\int f(x)\, S(x,K)\, d^4x}{\int S(x,K)\, d^4x}\, ,\\
\tilde \cx^\mu & = & \cx^\mu - \langle \cx^\mu \rangle \, .
\end{eqnarray}
\end{subequations}
The explicit azimuthal dependence is  displayed in eqs.~\eqref{mix}.
In addition, the (co-)variances 
$\langle\tilde \cx^\mu \tilde\cx^\nu \rangle$ can
depend on $\phi$ and this is the implicit azimuthal dependence. 
From eqs.~\eqref{mix}, correlation radii can be calculated for both 
Models just by inserting the corresponding emission function.

Azimuthal dependence of the correlation radii can be analysed with the
help of Fourier decomposition. Due to a number of symmetry arguments
\cite{Heinz:2002au} in our setup, the relevant Fourier series,
truncated after the leading oscillating terms, are
\begin{subequations}
\label{FDhbt}
\begin{eqnarray}
R_o^2(\phi) & = & R_{o,0}^2 + 2 R_{o,2}^2 \cos 2\phi + \dots\\
R_s^2(\phi) & = & R_{s,0}^2 + 2 R_{s,2}^2 \cos 2\phi + \dots\\
R_{os}^2(\phi) & = & 2 R_{os,2}^2 \sin 2\phi + \dots\\
R_l^2(\phi) & = & R_{l,0}^2 + 2 R_{l,2}^2 \cos 2\phi + \dots \, . 
\end{eqnarray}
\end{subequations}
%


\subsection{Results}
\label{hbt:results}

We will focus on $R_o^2$ and $R_s^2$, as we are interested in 
anisotropies in the transverse plane. The absolute sizes of these 
radii together with their oscillation amplitudes scale with the total 
geometric size $R$. We can get rid of this scaling and thus 
observe the effect due to anisotropies more cleanly when we study 
the ratios $R_{o,2}^2/R_{o,0}^2$ and 
$R_{s,2}^2/R_{s,0}^2$ \cite{retiere}.
%
\begin{figure}[h]
\begin{center}\begin{minipage}{8.3cm}
\epsfig{file=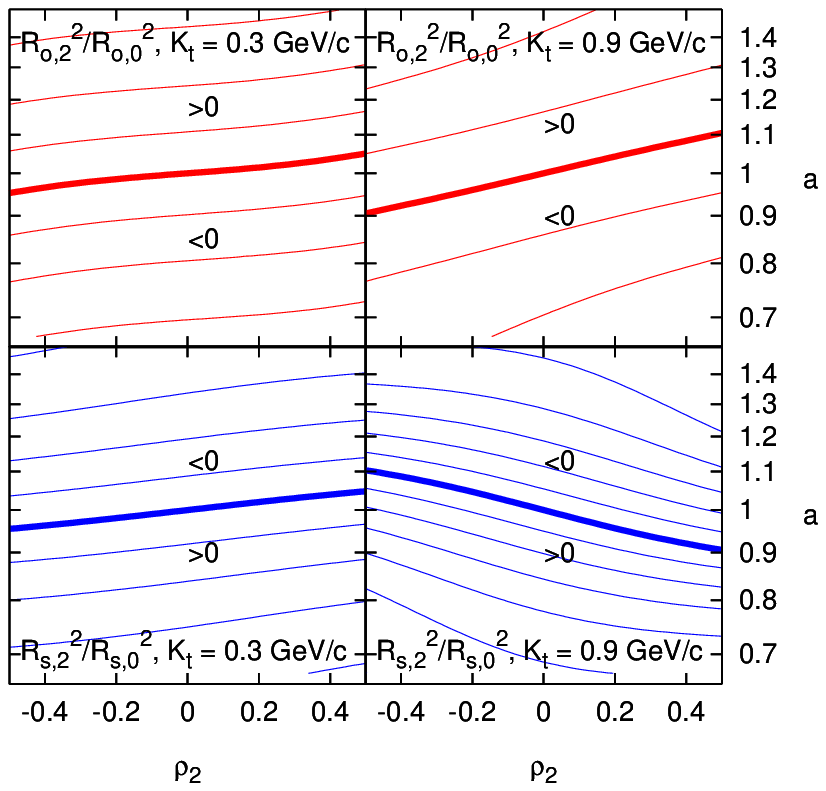,width=8.3cm}%
\vspace*{-0.9cm}
\centerline{Model 1}
\centerline{}
\end{minipage}
\begin{minipage}{8.3cm}
\epsfig{file=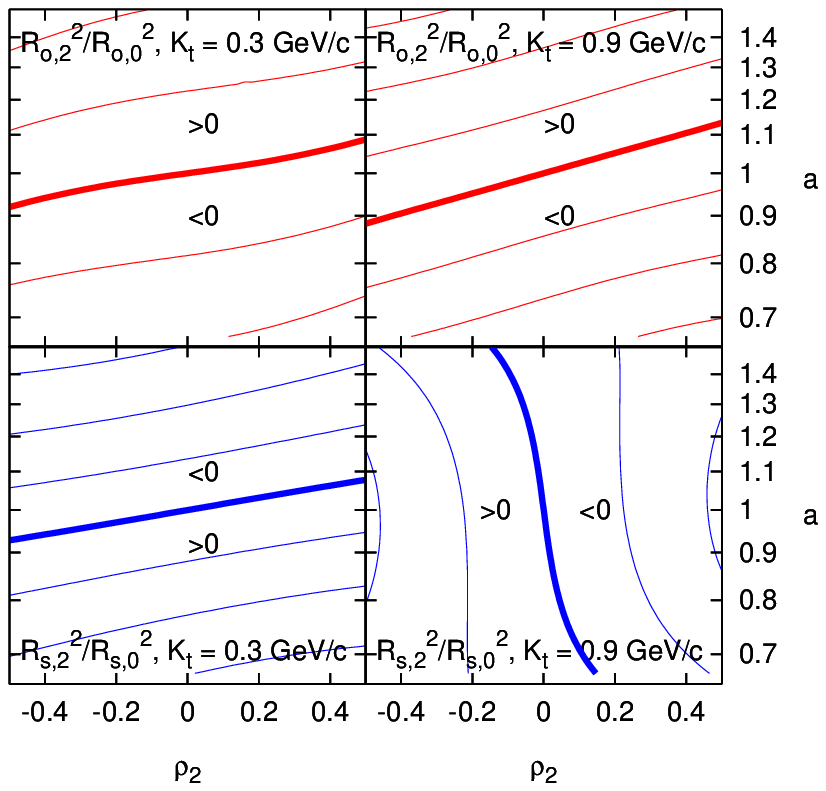,width=8.3cm}%
\vspace*{-0.9cm}
\centerline{Model 2}
\centerline{}
\end{minipage}
\end{center}
\caption{%
Dependence of the normalised second order oscillation terms 
$R_{o,2}^2/R_{o,0}^2$ (upper rows) and $R_{s,2}^2/R_{s,0}^2$ 
(lower rows) on spatial anisotropy $a$ and flow anisotropy 
$\rho_2$, calculated for $K_t = 300\, \mbox{MeV}/c$ (left columns)
and $K_t = 900 \, \mbox{MeV}/c$ (right columns), with Model 1
(upper panel) and Model 2 (lower panel). Thick contour lines correspond
to 0, consecutive curves to increments/decrements by 0.1. Other 
model parameters in the calculation were $T=0.1\, \mbox{GeV}$,
$\rho_0 = 0.88$, $R = 9.41\, \mbox{fm}$, $\tau_0 = 9\, \mbox{fm}/c$,
and $\Delta\tau = 1\, \mbox{fm}/c$. 
\label{f:hm12}}
\end{figure}
%

These ratios for Model 1 and Model 2 are plotted in Figure~\ref{f:hm12}.
In most cases, oscillations of correlation radii are mainly determined
by the spatial anisotropy and not so much by the flow anisotropy.
The only exception is $R_s^2$ at high $\pt$ in Model 2: the 
$\phi$-dependence in this model changes from shape-determined to 
flow-determined, i.e.\ dominated by the {\em implicit} azimuthal dependence.
In all other cases, $\phi$-dependence of the correlation radii 
follows the explicit azimuthal angle dependence rather well.

%
\begin{figure}
\centerline{\epsfig{file=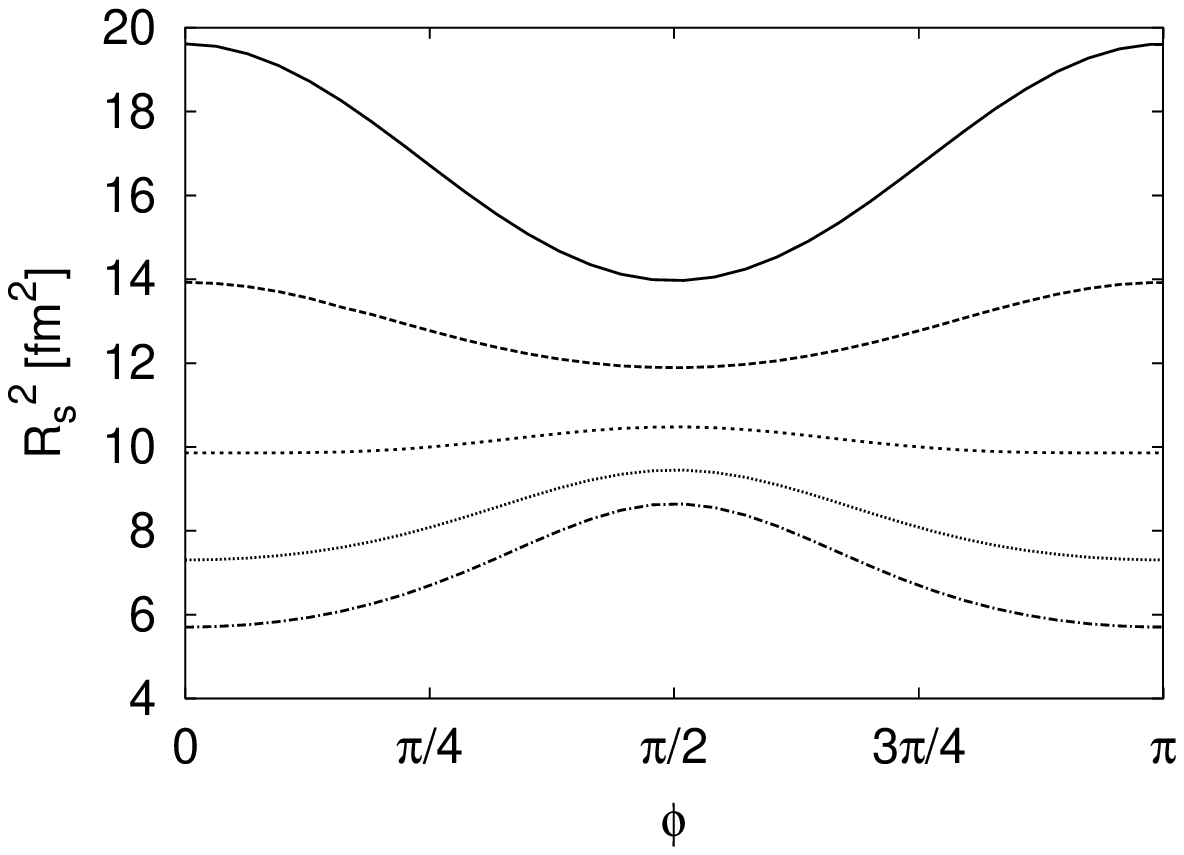,width=9.1cm}}%
\caption{%
Azimuthal angle dependence of $R_s^2$ in Model 2 at various 
transverse momenta. Values of parameters used in the calculation:
$T = 0.1\, \mbox{GeV}$, $\rho_0 = 0.88$, $\rho_2 = 0.2$,
$R = 9.41\, \mbox{fm}$, $a = 0.95$, $\tau_0 = 9\, \mbox{fm}/c$,
and $\Delta\tau = 1\, \mbox{fm}/c$. Different curves correspond 
from top to bottom to transverse momenta of 0.2, 0.4, 0.6, 0.8, and
1 GeV/$c$.
\label{f:Rs2}}
\end{figure}
%
The behaviour of $R_s^2$ in {\em Model 2} is shown in Figure~\ref{f:Rs2}. 
The azimuthal angle dependence changes 
{\em qualitatively} with increasing $\pt$:
at the given parameter values oscillation amplitude is positive 
at low $\pt$ and becomes negative at large $\pt$.

%
\begin{figure}[h]
\begin{center}
\begin{minipage}{9.9cm}
\centerline{\epsfig{file=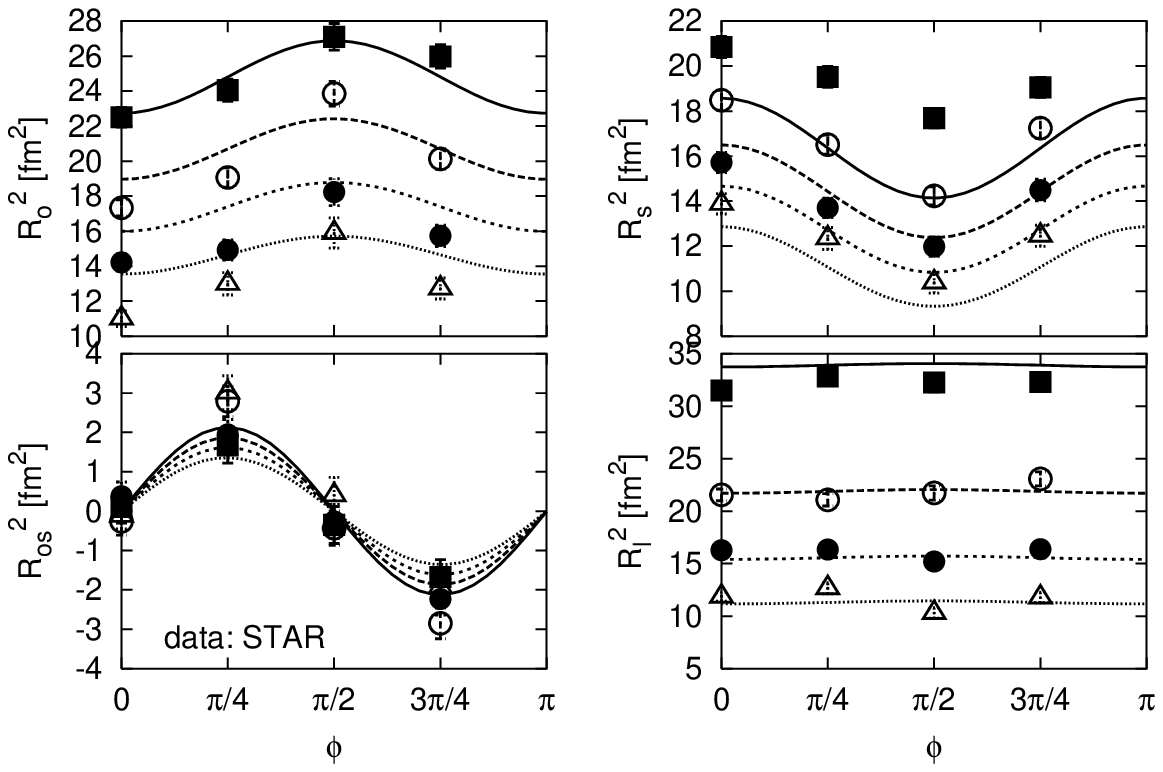,width=9.6cm}}%
\centerline{Model 1}
\centerline{}
\end{minipage}
\begin{minipage}{9.9cm}
\centerline{\epsfig{file=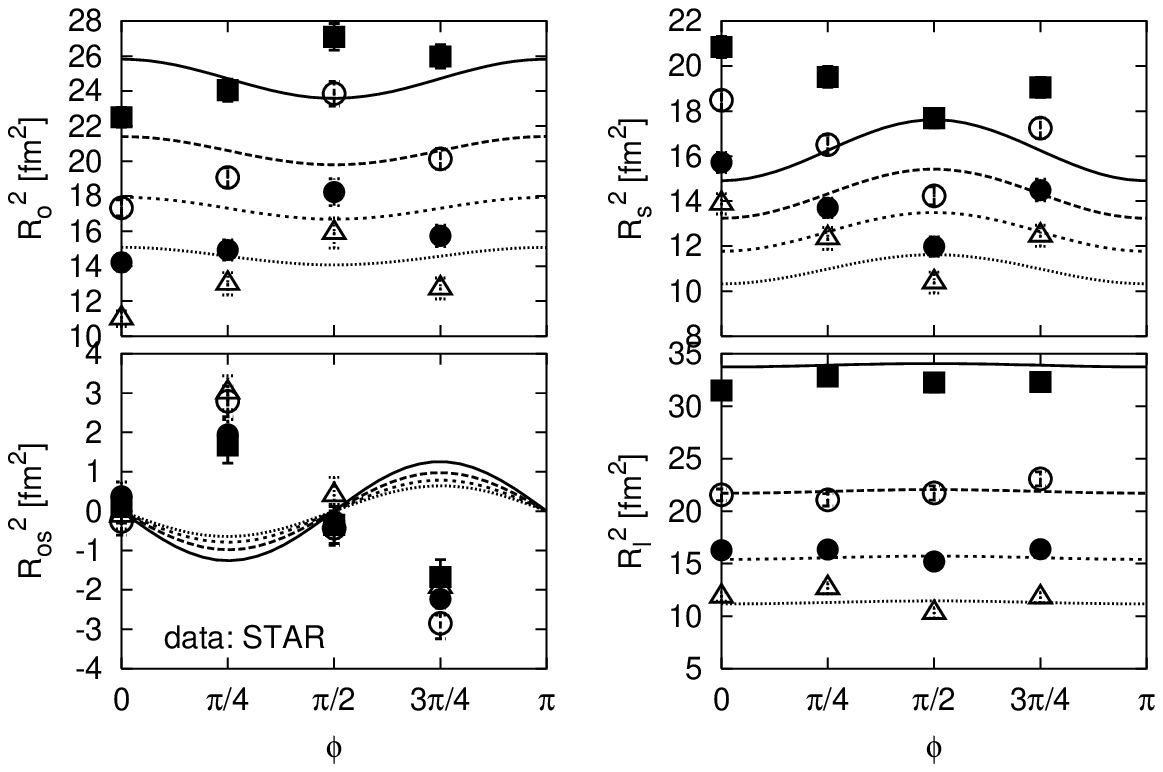,width=9.6cm}}%
\centerline{Model 2}
\centerline{}
\end{minipage}
\end{center}
\caption{%
Azimuthal angle dependence of correlation radii in Model 1 (left)
and Model 2 (right) in comparison with data. Curves and data points 
correspond from top
to bottom to transverse momenta of 0.2, 0.3, 0.4, and 0.52~GeV/$c$. 
Values of parameters used in the calculation are chosen such that 
$v_2(\pt)$ is reproduced (not shown here, see \cite{procpap}):
$T = 0.12\, \mbox{GeV}$, $\rho_0 = 0.99$, $\rho_2 = 0.035$,
$R = 9.41\, \mbox{fm}$, $\tau_0 = 5.02\, \mbox{fm}/c$,
and $\Delta\tau = 2.90\, \mbox{fm}/c$. Spatial anisotropies
are  $a = 0.946$ (Model 1) or $a = 1.057$ (Model 2). Data points are measured 
by the STAR collaboration \cite{stAHBT} in 20-30\% centrality events
of Au+Au collisions at $\sqrt{s}=200\, A\mbox{GeV}$.
\label{f:datacomp}}
\end{figure}
%
Now we come to the question if one of the models can be disqualified 
by comparing to data. In Figure~\ref{f:datacomp} we see two models
which reproduce $v_2$ quite well (this is shown in \cite{procpap}); 
they are related by transformation $a\to a^{-1}$. However, since 
the oscillation of correlation radii is mostly shaped by the 
spatial anisotropy, the two models lead to opposite predictions
of the sign of the oscillation amplitudes---and Model 2
is in qualitative disagreement with the data. I do not try 
to find the perfect fit here; this turns out to be a problematic
task with the blast-wave model \cite{fabproc,procpap}. Nonetheless,
note that the {\em qualitative} features of the 
azimuthal angle dependence of the correlation radii are reproduced
in Model 1 under the assumption of an {\em out-of-plane} extended source,
which confirms the earlier conclusion of STAR \cite{stAHBT}

\section{Conclusions}

Generalisation of the popular blast-wave model to non-central 
collisions is not unique. Many possible ways differ in how the 
transverse velocity depends on the azimuthal angle $\phi_s$. 

In this paper two such  generalisations were constructed.
Then, a number of statements scattered in literature were demonstrated 
in a unified framework of a generalised blast-wave model, which is often 
used in many variations. The interplay of HBT analysis with $v_2$
for identified species should be stressed.

I showed analytically how two very different fireballs can lead to the same 
$v_2$, such that from measuring only this quantity one cannot
conclude whether the source is elongated in-plane or out-of-plane
\cite{st130}.

The azimuthal angle dependence of the correlation radii, on the other 
hand, can be used for this. It is mostly sensitive at low $\pt$
to the spatial anisotropy of the fireball \cite{khplb}.

When the type of the model is identified from comparison to
data on two-pion correlations, spatial and flow anisotropy 
can be disentangled from reproducing {\em $v_2(\pt)$
of different identified species} and {\em $\phi$-dependence 
of correlation radii}. Before measuring anisotropies, temperature
and radial flow can be determined from azimuthally integrated 
single-particle spectra which are nearly independent of the 
anisotropy parameters.

Among the two models used here, Model 2 fails {\em qualitatively}
in reproducing data on azimuthal angle dependence of the correlation 
radii. In a simple qualitative comparison with the data, the 
other Model indicates that the observed fireball in non-central 
Au+Au collisions at $\sqrt{s}=200\, A\mbox{GeV}$ is elongated
out of the reaction plane, in agreement with conclusions
of \cite{stAHBT}.


\subsection*{Acknowledgements}
I thank Evgeni Kolomeitsev, Mike Lisa, Scott Pratt, and Fabrice Reti\`ere
for stimulating discussions. 

This research was supported by a Marie Curie Intra-European Fellowship
within the 6th European Community Framework Programme.


\appendix

\section{Jacobians for integration in transverse plane}
\label{sjaco}

As it is simpler, we begin with the Model 2.
The aim is to use $\rt$ and $\phs$ as coordinates in the transverse
plane. Because
\begin{subequations}
\begin{eqnarray}
\cx & = & r\, \cos\phs \\
\cy & = & r\, \sin\phs\, ,
\end{eqnarray}
\end{subequations}
we only have to replace $r$ by $\rt$. From eqs.~\eqref{rtil} and 
\eqref{adef} we obtain 
\begin{equation}
\rt = \frac{r}{R} \sqrt{a^{-2}\cos^2\phs + a^2\sin^2\phs}\, .
\end{equation}
This leads to 
\begin{subequations}
\label{xyps}
\begin{eqnarray}
\cx & = & \frac{\rt\, R\, \cos\phs}{\sqrt{a^{-2} \cos^2\phs + a^2\sin^2\phs}}\\
\cy & = & \frac{\rt\, R\, \sin\phs}{\sqrt{a^{-2} \cos^2\phs + a^2\sin^2\phs}}
\, ,
\end{eqnarray}
\end{subequations}
and 
\begin{equation}
\label{jam2}
d\cx\, d\cy = 
\frac{R^2\, \rt\, d\rt \, d\phs}{a^{-2}\cos^2\phs + a^2\sin^2\phs}\, .
\end{equation}
Thus we derived eq.~\eqref{jac2}.

For the Model 1 we want to use the angle $\phb$ as a coordinate instead of
$\phs$. By making use of eq.~\eqref{pbps} we can rewrite eq.~\eqref{xyps}
into
\begin{subequations}
\label{xypb}
\begin{eqnarray}
\nonumber
\cx & = & 
\frac{\rt\, R\, \mbox{sgn}(\cos\phs)}{\sqrt{a^{-2}+a^2\tan^2\phs}} \\
& = &
\frac{\rt\, R\, a^2\, \cos\phb}{\sqrt{a^2\cos^2\phb+a^{-2}\sin^2\phb}} \\
\nonumber
\cy & = & 
\frac{\rt\, R\, \mbox{sgn}(\sin\phs)}{\sqrt{a^{-2}\mbox{cotan}^2\phs + a^2}}\\ 
& = &
\frac{\rt\, R\, a^{-2}\, \sin\phb}{\sqrt{a^2\cos^2\phb+a^{-2}\sin^2\phb}}\, .
\end{eqnarray}
\end{subequations}
In the last expression we introduced 
\[
\mbox{sgn}(x) = \Big \{ \begin{array}{r@{\quad:\quad}l}
                          + 1 & x\ge 0 \\
                          - 1 & x < 0
                         \end{array}  \, ,
\]
and exploited that
\begin{eqnarray}
\nonumber
\mbox{sgn}(\sin\phs) & = & \mbox{sgn}(\sin\phb) \\
\mbox{sgn}(\cos\phs) & = & \mbox{sgn}(\cos\phb) \, .
\nonumber
\end{eqnarray}
From eqs.~\eqref{xypb} it is straightforward to obtain the 
Jacobian for Model 1
\begin{equation}
\label{jam1}
d\cx\, d\cy = 
\frac{R^2\, \rt\, d\rt\, d\phb}{a^2 \cos^2\phb + a^{-2}\sin^2\phb}\,  .
\end{equation}
Notice that apart from the use of $\phb$ instead of $\phs$, one 
obtains the Jacobian for Model 2 from that of Model 1 just by
replacing $a \to a^{-1}$.


\section{Calculation of $v_2$}
\label{v2calc}

Let us calculate $v_2$ for Model 1. First, we need the azimuthally 
integrated single-particle spectrum in the denominator of 
eq.~\eqref{v2comp}. From eqs.~\eqref{pdec} and \eqref{um1} we obtain 
that 
\begin{equation}
\label{estar1}
p^\mu u_\mu = \mt \, \cosh(\eta - y)\, \cosh\rho(\rt,\phb)  -
\pt\, \sinh\rho(\rt,\phb)\, \cos(\phi-\phb)\, .
\end{equation}
This is the energy argument for the Boltzmann distribution. In 
accord with eqs.~\eqref{ps} and \eqref{efun}, azimuthally integrated
spectrum is obtained as
\begin{multline}
\int_0^{2\pi} P_1(\pt,\phi)\, d\phi \\
\shoveleft{\, =  \int_0^{2\pi} d\phi \int d^4x\, S(x,p)}\\
\shoveleft{\, = \int_0^{2\pi} d\phi \int_0^{2\pi} d\phb\, J_1(\phb) 
\int_0^1 d\rt\, \rt\, R^2 \, 
\int_{-\infty}^{\infty} d\eta\, \frac{\mt\cosh(\eta-y)}{(2\pi)^3}}\\
\shoveleft{\quad \times 
\int_{-\infty}^{\infty} \frac{d\tau\, \tau}{\sqrt{2\pi\, \Delta\tau^2}}
\, \exp \left ( - \frac{(\tau - \tau_0)^2}{2\, \Delta\tau^2} \right ) }
\\  
\quad \times \exp \left ( - \frac%
{\mt \, \cosh(\eta-y)\, \cosh\rho(\rt,\phb) 
      - \pt\, \sinh\rho(\rt,\phb)\, \cos(\phi-\phb)}%
{T} \right ) 
\end{multline}
where $J_1(\phb)$ was defined in eq.~\eqref{jac1}.
The integration over $\tau$ is trivial. We are interested in 
mid-rapidity particles in the centre-of-mass frame, so $y=0$.
Then, integration over $\eta$ can be performed and leads to the 
modified Bessel function $\bkj$ \cite{abram}. We can exchange 
the order of integrations in $\phi$ and $\phb$, and perform a 
transformation $\phi\to \phi-\phb = \psi$. The integral in $\psi$
can then be performed analytically and leads to the modified Bessel function 
$\bin$. We finally arrive at 
\begin{eqnarray}
\label{den1}
\int_0^{2\pi} P_1(\pt,\phi)\, d\phi & = &
\frac{R^2\, \tau_0\, \mt}{2\pi^2} \int_0^1 d\rt\, \rt 
\int_0^{2\pi} d\phb\, J_1(\phb) 
\\ && \times
\bkj\left(\frac{\mt\, \cosh\rho(\rt,\phb)}{T}\right )\,
\bin\left ( \frac{\pt\, \sinh\rho(\rt,\phb)}{T} \right ) \, .
\nonumber
\end{eqnarray}

The numerator of eq.~\eqref{v21} is obtained in a similar way 
as the azimuthally integrated spectrum, we just add a factor 
$\cos(2\phi)$. After performing the integration over $\tau$
and $\eta$ we obtain 
\begin{multline}
\int_0^{2\pi} P_1(\pt,\phi)\, \cos(2\phi)\, d\phi \\
\shoveleft{ \quad = \frac{R^2\, \tau_0\, \mt}{4\pi^3} \int_0^1 d\rt\, \rt 
\int_0^{2\pi} d\phb\, J_1(\phb) \, 
     \bkj\left( \frac{\mt\, \cosh\rho(\rt,\phb)}{T} \right )} \\
\times \int_0^{2\pi} d\phi\, \cos(2\phi) \, 
\exp \left ( \frac{\pt\, \sinh\rho(\rt,\phb)}{T} \cos(\phi-\phb)\right )\, .
\end{multline}
Now again, we write $\phi=\psi+\phb$ and decompose
\begin{eqnarray*}
\cos(2\phi) & = & \cos(2\psi+2\phb)\\
& = &
\cos(2\psi)\cos(2\phb)-\sin(2\psi)\sin(2\phb)\, .
\end{eqnarray*}
The $\psi$-integral with the term proportional to $\sin(2\psi)$ vanishes.
The second term, proportional to $\cos(2\psi)\exp(\# \cos\psi)$ leads
to a modified Bessel function $\bid$ \cite{abram}. As a result we 
thus obtain
\begin{multline}
\label{num1}
\int_0^{2\pi} P_1(\pt,\phi)\, \cos(2\phi)\, d\phi \\
\shoveleft{ \quad = \frac{R^2\, \tau_0\, \mt}{2\pi^2}\, \int_0^1 d\rt\, \rt 
\int_0^{2\pi} d\phb\, J_1(\phb)\, \cos(2\phb) }\\
\times \bkj\left (\frac{\mt\cosh\rho(\rt,\phb)}{T}\right )\,
\bid\left (\frac{\pt \, \sinh\rho(\rt,\phb)}{T} \right )\, .
\end{multline}
By dividing this equation with the denominator derived in eq.~\eqref{den1}
we obtain the expression \eqref{v21} for $v_2$.

Calculation for Model 2 follows exactly the same steps as we've gone 
with Model 1. The only difference is that $\phb$ is replaced 
by $\phs$ and one uses $J_2(\phs)$ instead of $J_1(\phb)$.



\begin{thebibliography}{99}

\bibitem{vol}
S.~Voloshin and Y.~Zhang, Z.~Phys. C \textbf{70}, 665 (1996).

\bibitem{phv2}
S.S.~Adler \textit{et al.} [PHENIX Collaboration], 
Phys.~Rev.~Lett.~\textbf{91} 182301 (2003).

\bibitem{stv2}
J.~Adams \textit{et al.} [STAR Collaboration],
{nucl-ex/0409033}.

\bibitem{Heinz:2001xi}
U.~Heinz and P.F.~Kolb,
Nucl.\ Phys.\ A {\bf 702}, 269 (2002)
[arXiv:hep-ph/0111075].

\bibitem{Molnar:2001ux}
D.~Moln\'ar and M.~Gyulassy,
Nucl.\ Phys.\ A {\bf 697}, 495 (2002)
[Erratum-ibid.\ A {\bf 703}, 893 (2002)]
[arXiv:nucl-th/0104073].

\bibitem{st130}
C.~Adler {\it at al.} [STAR Collaboration],
Phys.~Rev.~Lett.\ {\bf 87}, 182301 (2001). 

\bibitem{Wiedemann:1997cr}
U.A.~Wiedemann,
Phys.\ Rev.\ C \textbf{57}, 266 (1998).

\bibitem{khplb}
U.~Heinz and P.F.~Kolb, Phys.~Lett.~B~{\bf 542}, 216 (2002).

\bibitem{Csorgo:1995bi}
T.~Cs\"org\H o and B.~L\"orstad,
Phys.\ Rev.\ C {\bf 54}, 1390 (1996).


\bibitem{retiere}
F.~Reti\`ere and M.A.~Lisa, 
{nucl-th/0312024}.

\bibitem{cf}
F.~Cooper and G.~Frye,
Phys.\ Rev.\ D \textbf{10} 186 (1974).

\bibitem{bjork}
J.~D.~Bjorken,
Phys.\ Rev.\ D \textbf{27}, 140 (1983).

\bibitem{Makhlin:1987gm}
A.~N.~Makhlin and Y.~M.~Sinyukov,
Z.\ Phys.\ C \textbf{39}, 69 (1988).

\bibitem{huo}
P.~Huovinen \textit{et al.}, Phys.~Lett.~B \textbf{503}, 58 (2001).

\bibitem{Tomasik:2002rx}
B.~Tom\'a\v sik and U.~A.~Wiedemann,
in Quark Gluon Plasma 3, R.C.~Hwa and X.-N.~Wang eds.,
World Scientific, 2003, pp. 715-777, 
[{hep-ph/0210250}].

\bibitem{Heinz:2002au}
U.~W.~Heinz, A.~Hummel, M.~A.~Lisa and U.~A.~Wiedemann,
Phys.\ Rev.\ C \textbf{66}, 044903 (2002).

\bibitem{Lisa:2000xj}
M.~A.~Lisa \textit{et al.}  [E895 Collaboration],
Phys.\ Lett.\ B \textbf{496}, 1 (2000).

\bibitem{procpap}
B.~Tom\'a\v sik, proceedings of 18~Conference of Nuclear Physics Division 
of the EPS, Prague, Aug. 23--29, 2004, to be published in Nucl. Phys. A,
{nucl-th/0409075}.

\bibitem{stAHBT}
J.~Adams \textit{et al.} [STAR Collaboration],
Phys. Rev. Lett. \textbf{93}, 012301 (2004).

\bibitem{fabproc}
F.~Reti\`ere, J.~Phys.~G \textbf{30}, S827 (2004).

\bibitem{abram}
M.~Abramowitz and I.A.~Stegun, Handbook of Mathematical Functions, Dover,
New York, 1964--1972.

\end{thebibliography}
\end{document}